\documentclass[11pt,a4paper]{article}
\pdfoutput=1
\usepackage{jcappub}
\usepackage{graphicx}
\usepackage{longtable,comment}
\usepackage{amsmath}
\usepackage{amssymb}
\usepackage{subfigure}
\usepackage{hyperref}
\usepackage{dcolumn}
\usepackage{multirow}
\usepackage{cancel}
\usepackage[normalem]{ulem}
\newcommand{\rout}{\bgroup\markoverwith
{\textcolor{red}{\rule[.5ex]{2pt}{0.4pt}}}\ULon}
\usepackage{cancel}
\usepackage{xcolor}

\usepackage{color}

\newcommand{\be}{\begin{eqnarray}}
\newcommand{\ee}{\end{eqnarray}}
\newcommand{\mpl}{{M_{\rm {pl}}}}

\newcommand{\dd}{\, {\rm d}}
\newcommand{\gsim}{\;\mbox{\raisebox{-0.5ex}{$\stackrel{>}{\scriptstyle{\sim}}$}
}\;}
\newcommand{\lsim}{\;\mbox{\raisebox{-0.5ex}{$\stackrel{<}{\scriptstyle{\sim}}$}
}\;}
\def\eea{\end{eqnarray}}
\def\bea{\begin{eqnarray}}
\newcommand{\ppp}{^\prime}
\newcommand{\tg}{\tilde{g}}

\newcommand{\nm}{{\mu\nu}}
\newcommand{\ppn}{{\rm PPN}}

\newcommand{\mmm}{{_{\rm m}}}
\newcommand{\GN}{G_{\rm N}}

\newcommand{\ph}{\phi_0}
\newcommand{\phh}{\phi_1}
\newcommand{\phhh}{\phi_2}
\newcommand{\pd}{\partial}
\newcommand{\dph}{\dot{\phi}_0}
\newcommand{\dphh}{\dot{\phi}_1}
\newcommand{\ddphh}{\ddot{\phi}_1}

\newcommand{\oo}{\mathcal{O}}
\newcommand{\rs}{\rho}
\newcommand{\ddph}{\ddot{\phi}_0}

\newcommand{\tn}{\textnormal}

\newcommand{\upo}{\Upsilon}
\newcommand{\upt}{\Sigma}
\newcommand{\bpp}{\beta_\phi}
\newcommand{\gti}{\tilde{\gamma}}
\newcommand{\Tmm}{{T_{{\rm m}\,\mu\nu}}}
\newcommand{\s}{\textrm{SS}}

\allowdisplaybreaks

\newcommand{\refeq}[1]{Eq.~(\ref{eq:#1})}          
\newcommand{\refeqs}[2]{Eqs.~(\ref{eq:#1})--(\ref{eq:#2})}

\begin{document}

\title{Solar System Constraints on Disformal Gravity Theories}

\author[a]{Hiu Yan Ip,} 

\author[b]{Jeremy Sakstein,}

\author[a]{Fabian Schmidt}

\affiliation[a]{Max-Planck-Institut f\"ur Astrophysik, Karl-Schwarzschild-Str. 1, 85741 Garching, Germany}

 \affiliation[b]{Institute of Cosmology and Gravitation,
University of Portsmouth, Portsmouth PO1 3FX, UK}

 \emailAdd{iphys@mpa-garching.mpg.de}
 \emailAdd{jeremy.sakstein@port.ac.uk}
 \emailAdd{fabians@mpa-garching.mpg.de}

\abstract{ Disformal theories of gravity are scalar-tensor theories where the scalar couples derivatively to matter via the Jordan 
frame metric. These models have recently attracted interest in the cosmological 
context since they admit accelerating solutions. We derive the solution for a static isolated mass in generic disformal gravity theories and transform 
it into the parameterised post-Newtonian form. This 
allows us to investigate constraints placed on such theories by local tests of gravity. The tightest constraints come from preferred-frame effects 
due to the motion of the Solar System with respect to the evolving cosmological background field.  
The constraints we obtain improve upon the previous solar system constraints by two orders
of magnitude, and constrain the scale of the disformal coupling for generic
models to $\mathcal{M} \gtrsim 100$~eV. These constraints render all disformal effects irrelevant for cosmology.}

\maketitle

\section{Introduction}
\label{sec:intro}

The acceleration of the cosmic expansion is one of the biggest mysteries in modern cosmology and theoretical physics. The search for the underlying 
driving mechanism, dubbed \textit{dark energy}, has previously prompted the study of light cosmological scalar fields as one potential 
candidate (see \cite{Copeland:2006wr,Clifton:2011jh,Joyce:2014kja} for reviews). Theories where a scalar couples to matter, scalar-tensor theories, 
can be seen as  modifications 
of general relativity (GR); these have been the subject of a considerable amount of recent research. Among the plethora of models in the literature, those that contain 
screening mechanisms \cite{Jain:2010ka,Khoury:2010xi} are particularly well-studied due to their ability to decouple solar system scales from 
cosmological ones. This circumvents the need for fine-tuning the model parameters to unnaturally small numbers, as the modifications can 
still be a dominant influence in the cosmological background while satisfying solar system bounds. Scalar-tensor theories are most commonly 
realised through a conformal 
coupling of the scalar $\phi$ to matter via a metric relation of the form 
\begin{equation}
\tg_{\mu\nu}=A^2(\phi)g_\nm\,,
\end{equation}
where $g_\nm$ obeys equations of motion derived from the Einstein-Hilbert action
(Einstein-frame metric), while $\tg_\nm$ is the physical metric whose geodesics free-falling matter follows (Jordan-frame metric). However, this is
not the most general relation one can 
write down. Bekenstein \cite{Bekenstein:1992pj} showed that the most general relation that preserves causality is 
\begin{equation} \label{eq:bek}
 \tg_\nm=C(\phi,X)g_\nm+D(\phi,X)\partial_\mu\phi\partial_\nu\phi\quad
\mbox{where}\quad X\equiv-\frac{1}{2}g^{\nm}\partial_\mu\phi\partial_\nu\phi.
\end{equation}
The addition of this coupling to any theory does not introduce a ghost by itself.\footnote{This is assuming that the signs of $C$ and $D$ are 
chosen appropriately.} 
When $C$ and $D$ depend on $\phi$ only the theory is a subset of the Horndeski class\footnote{Unless non-Horndeski terms are present in the action $S[g_\nm, \phi]$.} 
\cite{Horndeski:1974wa,Bettoni:2013diz}, and when $D$ depends on $\phi$ and $X$ the theory fits into the ``beyond Horndeski'' class 
\cite{Gleyzes:2014qga}. The latter case contains hidden constraints that render the equations of motion second-order \cite{Zumalacarregui:2013pma}. 
The term proportional to 
$D$ is known as the disformal coupling and, in contrast to the first term---the conformal coupling---it has only recently begun to be studied in 
detail 
\cite{Kaloper:2003yf,Koivisto:2008ak,Zumalacarregui:2010wj,Noller:2012sv,vandeBruck:2012vq,Zumalacarregui:2012us,vandeBruck:2013yxa,Brax:2013nsa,
Brax:2014vva,Brax:2014zba,Sakstein:2014isa,Sakstein:2014aca,vandeBruck:2015ida,Koivisto:2015mwa,Hagala:2015paa,Domenech:2015hka}. Until recently, most 
studies have focused on the 
cosmology of these theories and, unlike the conformal coupling 
\cite{Davis:2011qf,Jain:2012tn,Brax:2013uh,Sakstein:2013pda,Vikram:2014uza,Sakstein:2014nfa,Sakstein:2015oqa,Koyama:2015oma}, little attention has 
been given to astrophysical and solar system effects. 
Furthermore,
Ref.~\cite{ Koivisto:2012za} claimed that disformally coupled theories have their own
screening mechanism. Ref.~\cite{Sakstein:2014isa} has examined the local behaviour of these theories and found that no such screening mechanism 
exists.
He further found that local 
objects only source scalar gradients (and hence modifications of gravity) when the background cosmological scalar $\phi_0(t)$ is accounted for. In this 
case, the sourcing is proportional to $\dph$ 
and $\ddph$ and, while there are no non-linear screening mechanisms, the modifications can be screened if the cosmology is such that these 
time-derivatives are small. In this case, the effects on matter are screened everywhere, in contrast to non-linear mechanisms that only hide the 
scalar field locally.  
Note that the natural scale for the background evolution is the Hubble scale,
so that at the present time we expect $\dph \sim H_0 \phi_0,\,\ddph \sim H_0^2 \phi_0$ unless the model is fine-tuned. 

In this paper, we rigorously calculate solar system observables for a disformal theory with canonical scalar action. 
In particular, we calculate the parameters appearing in the 
parametrised post-Newtonian (PPN) metric. The PPN formalism \cite{Will:1972zz} is a general framework for testing alternate theories of gravity in 
the solar system. Deviations from GR are encoded in 10 parameters (one Newtonian and nine post-Newtonian) that appear multiplying various terms in 
the metric. Solar system experiments constrain these parameters either directly or in combinations (see \cite{Will:2008dya} for a review) and so one 
can constrain a wide class of theories using the same experimental data. 

 Calculating up to the first post-Newtonian order (2PN), we confirm that 
there is no screening mechanism, and that at all local effects are proportional to $\dph^2$.  Thus, the findings of Ref.~\cite{Sakstein:2014isa} hold to higher order as well.  
As a diffeomorphism-invariant theory derived from a Lagrangian, 
disformally coupled scalar-tensor theories are semi-conservative.  
The non-zero 
PPN parameters are the Eddington light bending parameter $\gamma$, the amount of non-linearity in the field equations $\beta$, and the two preferred -frame parameters $\alpha_1$ and $\alpha_2$. In scalar-tensor theories, the
evolution of the cosmological background field provides a preferred 4-vector,
singling out the cosmological (CMB) frame as preferred frame.  When the
scalar is coupled disformally to matter this leads to significant
non-zero $\alpha_1,\,\alpha_2$.  
All of the PPN parameters are proportional to the same quantity, which is a combination of the time-derivative of the 
cosmological scalar and the disformal coupling scale $\Lambda$. 
The tightest constraint comes from $\alpha_2$, which is constrained 
by local tests to be of order $10^{-7}$ or smaller. 
For generic (non-fine-tuned) models, this constrains the coupling scale
to $\Lambda/H_0 \gtrsim 3\times 10^3$. Equivalently, normalizing the
scalar field to the Planck mass, the coupling is constrained such that
$\mathcal{M} = (M_{\rm Pl} \Lambda)^{1/2} \gtrsim 10^2$~eV. 
This means that any observable
consequences of the disformal coupling in the cosmos are negligibly small. 
Moreover, it rules out the disformal coupling as a source of the cosmic
acceleration. 
Apart from extreme fine tuning to make $\dph,\,\ddph$ vanishingly small
today (which will most likely not allow for an accelerating solution without
vacuum energy), the only way to evade these constraints is to augment the model
with the well-known chameleon or Vainshtein screening mechanisms. 

The paper is set out as follows: We begin in section \ref{sec:disf} by introducing disformal gravity theories and give a brief introduction to the 
PPN formalism in section \ref{sec:PPN}. The calculation of the PPN parameters for the minimal disformal model, which is merely chosen for clarity of the presentation, is worked out in detail in section 
\ref{sec:mincase}.  
In section \ref{sec:constraints} this is used to place new constraints on $H_0/\Lambda$ and the first derivative of the 
disformal factor for a well-studied exponential model. The constraints on the disformal mass-scale $\mathcal{M}$ are two-orders of magnitude 
stronger than previous solar system constraints \cite{Sakstein:2014isa} but the constraints on the first derivative of the disformal factor are 
relatively weak and values as large as $10^3$ are not excluded. We conclude in section \ref{sec:concl}.  In Appendix~\ref{sec:gencalc} we show that additional contributions to the results in the 
most general disformal theory are sub-leading and hence our results of section \ref{sec:mincase} and \ref{sec:constraints}
hold for a much wider class of conformal/disformal models.

\section{Disformal Theories of Gravity}
\label{sec:disf}

Canonical disformal gravity theories are described by the scalar-tensor action:
\begin{equation}\label{eq:act}
 S=\int\dd^4 x \frac{\mpl^2}{2}\sqrt{-g}\left[\frac{\mpl^2 R(g)}{2}-\frac{1}{2}\nabla_\mu\phi\nabla^\mu\phi-V(\phi)\right]+S\mmm[\tg_\nm],
\end{equation}
which describes a massless spin-2 graviton and an additional scalar degree of freedom in the Einstein frame. The modifications of general relativity 
arise due to a coupling of the scalar to matter via the Jordan frame metric
\begin{equation}\label{eq:jfmet}
 \tg_\nm=A^2(\phi)\left[g_\nm+\frac{B^2(\phi)}{\Lambda^2}\partial_\mu\phi\partial_\nu\phi\right].
\end{equation}
The dimensionless function $A(\phi)$ describes the conformal coupling via the Einstein frame metric $g_\nm$, while the dimensionless function $B(\phi)$ controls the strength of the 
derivative interaction, which is known as the disformal coupling. Note that we have normalized $\phi$ to be dimensionless. In this paper, we will 
only study the case where the coupling functions depend solely on the scalar field, $\phi$. 
Our leading constraints will also apply to models that generalize this
to $B(\phi, X)$. The mass scale $\Lambda$ controls the amplitude of disformal effects, with smaller values of $\Lambda$ leading to larger effects.

In what follows we will write $\mpl^2=(8\pi G)^{-1}$ but it is important to note 
that $G$ is not necessarily equal to Newton's constant as measured by local experiments and, indeed, we will see later that it is not. The equations 
of motion in the Einstein frame are \cite{Sakstein:2014isa}
\begin{align}
 G_\nm=\:& 8\pi G\left(\Tmm+T_{\phi\,\nm}\right)\label{eq:EE}\\
 \left(1-\frac{2XB^2(\phi)}{\Lambda^2}\right) \Box\phi =\:& \frac{8\pi G B^2}{\Lambda^2} T_{\rm m}^\nm \nabla_\mu\nabla_\nu \phi 
-8\pi \alpha G T\mmm \nonumber\\
& -\frac{8\pi G B^2}{\Lambda^2} (\alpha_\phi(\phi)-\bpp(\phi))T_{\rm m}^\nm \pd_\mu\phi \pd_\nu \phi+V(\phi)\label{eq:phi_eom},
\end{align}
where $X=-1/2g^{\mu\nu}\partial_\mu\phi\partial_\nu\phi$ and we have defined\footnote{We define the derivative of the conformal and disformal factors as $\alpha_\phi,\,\bpp$ to 
avoid any confusion with the PPN parameters $\alpha_i,\,\beta$. The parameter $\beta_\phi$ was called $\gamma$ in \cite{Sakstein:2014isa,Sakstein:2014aca}.}
\begin{equation}\label{eq:agdefs}
 \alpha_\phi(\phi)\equiv\frac{\dd\ln A(\phi)}{\dd \phi}\quad \textrm{and}\quad\bpp(\phi)\equiv\frac{\dd\ln B(\phi)}{\dd \phi}.
\end{equation}
Here, $T_\mmm^\nm=\frac{2}{\sqrt{-g}}\frac{\delta S\mmm[g]}{\delta g_\nm}$ is the energy-momentum tensor of the matter fields defined with respect to the Einstein-frame metric and 
\begin{equation}\label{eq:emphi}
 T_{\phi\,\mu\nu}=\frac{1}{8\pi
G}\left[\nabla_\mu\phi\nabla_\nu\phi-g_\nm\left(\frac{1}{2}\nabla_\alpha\phi\nabla^\alpha\phi+V(\phi)\right)\right]
\end{equation}
is the energy-momentum tensor of the field. Note that these are not separately covariantly conserved quantities due to the coupling of the scalar to 
matter, only their sum is. Equation (\ref{eq:EE}) is simply Einstein's equation, which is a consequence of working in the Einstein frame. Equation 
(\ref{eq:phi_eom}) is the equation of motion for the scalar and this is where the modifications of gravity become apparent. Note that the physical 
metric is the Jordan frame metric (\ref{eq:jfmet}); it is this metric that governs the motion of test particles. The energy-momentum tensor of 
matter appears in equation (\ref{eq:phi_eom}) and so the field is sourced by any non-zero matter distribution characterised by $\Tmm$. In non-relativistic systems, 
the gravitational field is sourced mainly by matter and not the scalar, in which case the solution of (\ref{eq:EE}) is identical to the GR solution. 
The physical metric, and correspondingly the motion of test particles, then deviates from the GR prediction.

The goal of this paper is to calculate the physical (Jordan frame) metric to post-Newtonian order. The post-Newtonian predictions of purely conformal 
scalar-tensor theories have been well-studied (see \cite{will1993theory,Will:2008dya} and reference therein) but, to date, the post-Newtonian 
behaviour of disformal theories has yet to be derived. We therefore focus on the disformal part of (\ref{eq:jfmet}),
setting $A=1$ and $\alpha_\phi=0$.  
Furthermore, we set $B(\phi)$ to be constant ($\beta_\phi=0$), and absorb its value into $\Lambda$ to set $B=1$. As we 
will see below, the presence of a disformal coupling significantly complicates the calculation compared with the pure conformal case and so we will 
work with this minimal model first in order to make the calculation as simple as possible. In fact, in Appendix \ref{sec:gencalc} we show that the general model yields no new 
constraints compared with the minimal model. This is because the conformal
and disformal couplings are independently constrained by different
PPN parameters, so that any interaction of the two effects is highly suppressed. 
Ref.~\cite{Sakstein:2014isa} has examined the Newtonian behaviour 
of these theories and has shown that the disformal terms are only active when one accounts for the fact that the space-time is asymptotically FRW and 
not Minkowski and that the disformal terms are sourced by the time-derivative of the cosmological scalar $\phi_0$. The constraints for the non-minimal 
model are then essentially the same when setting $B \to B(\phi_0(t_0))$. 

The minimal model is described by the Jordan frame metric
\begin{equation}\label{eq:jfmin}
 \tg_\nm=g_\nm+\frac{\partial_\mu\phi\partial_\nu\phi}{\Lambda^2}\,.
\end{equation}
The inverse Jordan frame metric is given by
\begin{align}
\tg^\nm = g^\nm -\frac{1}{\Lambda^2}\left(1-\frac{2X}{\Lambda^2}\right)^{-\frac{1}{2}} g^{\mu \alpha}g^{\nu \beta} \pd_\alpha \phi \pd_\beta \phi\,.
\end{align}

\section{The PPN Formalism}
\label{sec:PPN}

With the exception of compact objects such as black holes and neutron stars, astrophysical objects move with 
non-relativistic velocities $v$ and one can solve Einstein's equations by expanding in appropriate powers of $v/c$. In what follows, we will work in units 
where $c=1$ and describe the expansion in powers of $1/c^{2m}$ as being $\oo(m)$ for brevity. In this sense, the solution of the 
equations to order $(v/c)^2$ is $\oo(1)$, the solution to order $(v/c)^3$ is $\oo(1.5)$ and the solution to order $(v/c)^4$ is $\oo(2)$. We will 
refer to the solutions at $\oo(1)$ as 1PN solutions and $\oo(2)$ as 2PN. 
Velocities $v$ are $\oo(0.5)$, while the Newtonian potential is
$\oo(1)$. Furthermore, time-derivatives add a power of $v/c$, while spatial derivatives do not. The leading effect in the disformal model comes from 
the time derivative of the background field $\dph$, which we take to be $\oo(0)$ (in the CMB rest frame).

The PPN framework parametrises the solution of Einstein's equations, or more generally the Jordan-frame metric, to 2PN 
as
\begin{eqnarray}
      \tg_{00} & = & - 1 + 2 U - 2 \beta U^2 - 2 \xi \Phi_W +\nonumber
      (2 \gamma + 2 + \alpha_3 + \zeta_1 - 2 \xi ) \Phi_1 +
      2 (3 \gamma - 2 \beta + 1 + \zeta_2 + \xi) \Phi_2 \\\nonumber
      & & + 2 (1 + \zeta_3) \Phi_3 + 2 (3 \gamma + 3 \zeta_4 - 2 \xi)
      \Phi_4 - (\zeta_1 - 2 \xi ) {\mathcal A} -
      (\alpha_1 - \alpha_2 - \alpha_3) w^2 U -
      \alpha_2 w^i w^j U_{ij}\label{eq:g00ppn} \\
      & & + (2 \alpha_3 - \alpha_1 ) w^i V_i,
      \\ 
      \tg_{0i} & = & - \frac{1}{2} (4 \gamma + 3 + \alpha_1 -\nonumber
      \alpha_2 + \zeta_1 - 2 \xi ) V_i - \frac{1}{2}
      (1 + \alpha_2 - \zeta_1 + 2 \xi) W_i -
      \frac{1}{2} ( \alpha_1 - 2 \alpha_2 ) w^i U \\\label{eq:g0ippn}
      & & - \alpha_2 w^j U_{ij},
      \\ 
      \tg_{ij} & = & (1 + 2 \gamma U) \delta_{ij},\label{eq:gijppn}
\end{eqnarray}%
where the 1PN potentials are 
\begin{equation}\label{eq:Udef}
 U\equiv \GN\int \dd^3 \vec{x}\ppp\frac{\rho (t, \vec{x}^\prime)}{|\vec{x}-\vec{x}\ppp|},\quad \tn{and}\quad 
 U_{ij}\equiv \GN\int \dd^3 \vec{x}\ppp\frac{\rho (t, \vec{x}^\prime) (x-x')_i(x-x')_j}{|\vec{x}-\vec{x}\ppp|^3}
\end{equation}
and the 2PN potentials are
\begin{align}
 \Phi_1&\equiv  \GN\int\dd^3 \vec{x}\ppp 
\frac{\rs(\vec{x}\ppp)v^2(\vec{x}\ppp)}{|\vec{x}-\vec{x}\ppp|}  ,\quad\Phi_2\equiv  \GN\int\dd^3 \vec{x}\ppp 
\frac{\rs(\vec{x}\ppp)U(\vec{x}\ppp)}{|\vec{x}-\vec{x}\ppp|}, \nonumber \\
\Phi_3&\equiv  \GN\int\dd^3 \vec{x}\ppp \frac{\rs(\vec{x}\ppp)\Pi(\vec{x}\ppp)}{|\vec{x}-\vec{x}\ppp|},  \quad \Phi_4\equiv  \GN\int\dd^3 \vec{x}\ppp 
\frac{p(\vec{x}\ppp)}{|\vec{x}-\vec{x}\ppp|},\nonumber\\  V_i& \equiv \GN\int\dd^3 \vec{x}\ppp 
\frac{v_i(\vec{x}\ppp)\rs(\vec{x}\ppp)}{|\vec{x}-\vec{x}\ppp|}, \quad W_i\equiv \GN\int\dd^3 \vec{x}\ppp \frac{ 
\rs(\vec{x}\ppp)\vec{v}\cdot(\vec{x}-\vec{x}\ppp)(x-x\ppp)_{ i}}{|\vec{x}-\vec{x}\ppp|^{ 3}}, \nonumber\\ \Phi_W&\equiv\GN^2\int\dd^3 x\ppp\dd^3 
x^{\prime\prime}\frac{\rho(x\ppp)\rho(x^{\prime\prime})(\vec{x}-\vec{x}^\prime)}{\left\vert\vec{x}-\vec{x}^\prime\right\vert^3}\cdot\left(\frac{\vec{x
} \ppp
-\vec{x}^{\prime\prime}}{\left\vert\vec{x}-\vec{x}^{\prime\prime}\right\vert}-\frac{\vec{x}
-\vec{x}^{\prime\prime}}{\left\vert\vec{x}\ppp-\vec{x}^{\prime\prime}\right\vert} \right)\quad\textrm{and}\nonumber \\ 
\mathcal{A}&\equiv\GN\int\dd^3x\ppp\frac{\rho(x\ppp)\left[\vec{v}(x\ppp)\cdot\left(\vec{x}-\vec{x}\ppp\right)\right]^2}{\left\vert\vec{x}-\vec{x}
\ppp\right\vert^3 }.
\label{eq:2PNpot}
\end{align}
Note that $\rho(t,\vec{x})$ is the rest-frame mass density of the fluid, which is important when transforming from a given set of coordinates to PPN.   
The PPN potentials are defined with respect to the locally measured Newton's constant $\GN$, which is not necessarily equal to the constant $G$ 
appearing in the action.  Instead, $\GN$ is defined such that it brings $g_{00}$ at 1PN order into the simple form $-1+2U$. The vector $w^i$, taken to 
be $\oo(0.5)$, is the speed of the PPN coordinate system relative to the mean rest frame of the universe and is taken to be constant in space and 
time, since it varies over time scales much longer than Solar System time scales. The PPN metric is not just a 
parametrisation of the solution but also a gauge choice: the coordinate time is equal to the proper time for an observer and the metric components 
scale as
\begin{align}
 \tg^{\rm PPN}_{00} &= \mathcal{O}(1) +\mathcal{O}(2),\\
 \tg^\ppn_{0i}&=\mathcal{O}(1.5)\quad\textrm{and}\\
 \tg^\ppn_{ij} & =\mathcal{O}(1).
\end{align}
In the following, we will refer to a tensor obeying this scaling as expanded to $\oo(2)$, since any scalar constructed out of it will be at least 
$\oo(2)$. 
The 10 parameters $\gamma$, $\beta$, $\xi$, $\zeta_i$ and $\alpha_i$ are arbitrary constants whose value depends on the 
specific theory in question. GR has $\gamma=\beta=1$ and $\xi=\zeta_i=\alpha_i=0$. Canonical conformally-coupled scalar-tensor theories on the other 
hand typically lead to different values of $\gamma$ and $\beta$ while the other PPN parameters remain at their vanishing GR values. The PPN 
parameters have been measured using a variety of different 
probes \cite{Will:2008dya} and are all consistent with GR. One can then calculate them in alternate theories and use the bounds to constrain the 
model 
parameters. We will do precisely this below. 

\section{Field solution and PPN parameters}
\label{sec:mincase}
Here we will calculate the PPN parameters for the theory described by the Jordan frame metric (\ref{eq:jfmin}). We will treat the scalar as a light, 
cosmological scalar driving the acceleration of the cosmic expansion. The
potential $V(\phi)$ and its derivatives are then all of order $H_0$, and we can neglect the mass of the scalar on the scales of interest. Screening 
mechanisms such as
the chameleon and symmetron can render the field locally massive
and thus hide it from local observations. In that case however, there is nothing
to add to the standard conformally coupled screened theories in the PPN context, and we will
not consider this case here.

\subsection{Preliminary Considerations}

\subsubsection{Choice of Coordinates}

We begin by setting up a coordinate system in which to solve the equations. 
We will derive
the field solution in the Einstein frame since the equations of motion are simplest there. The Einstein frame solution at zeroth-order in the CMB 
rest-frame is simply Minkowski space \footnote{We ignore corrections coming from the FRW nature of the zeroth-order solution since 
these are negligible on solar system scales.} i.e. the vacuum solution of the field equations:
\begin{equation}
 \dd s^2 = -\dd t^2 + \delta_{ij}\dd x^i \dd x^j.
\end{equation}
Note, however, that the Jordan frame metric {is
\begin{equation}
 \dd \tilde{s}^2 = -N^2\dd t^2 + \delta_{ij}\dd x^i\dd x^j\,,
\label{eq:JFmetric0}
\end{equation}
where the lapse is
\begin{equation}
N^2 = 1 - \frac{\dph^2}{\Lambda^2}\,.
\end{equation}
This differs from unity due to the presence of a time-dependent cosmological field $\phi_0$. In this sense, the proper time for a physical observer 
is not coincident with the coordinate time and so it is 
clear that even at lowest order the Jordan frame metric is not in the PPN gauge with this choice of coordinates. 
Note also that the Jordan frame 
energy-momentum tensor, $2/\sqrt{-\tg}\delta S\mmm/\delta \tg_\nm$ is related to the Einstein frame tensor introduced in the previous section by 
\cite{Zumalacarregui:2012us,Sakstein:2014isa}:
\begin{equation}\label{eq:emtrel}
T^{\nm}=\sqrt{1-\frac{2X}{\Lambda^2}}\tilde{T}^\nm.
\end{equation}
We will consider the metric sourced by a static object of finite extent and will hence treat its internal structure using a fluid description. Since the Jordan-frame
$\tilde{T}^\nm$ (rather than $T^\nm$) is the covariantly conserved energy-momentum tensor, all fluid variables such as the density and pressure must be defined in this 
frame.  We will hence consider the Einstein frame as a calculational tool; we will not assign any physical meaning to $T^\nm$, it is simply a source 
in the field equations. We will transform quantities to the Jordan-frame PPN gauge once the solutions at 1PN and 2PN have been found. 

\subsubsection{Preferred Frame Effects}

Our final goal is to derive the PPN metric in the rest frame of the Solar System, since that is the frame
in which observations are being made. There are two possible ways to proceed. First, one could directly derive the solution for the field in the
Solar System frame. In this frame, the background field gradient, which
is simply $\partial_\mu\phi_0= \dph (1,0,0,0)$
in the CMB frame, 
is Lorentz-boosted to
\begin{align}
(\pd_0 \ph)_\s =\:& \dph \left(1+\frac{w^2}{2}+\frac{3 w^4}{8}\right) \nonumber\\
(\pd_i \ph)_\s =\:& -\dph \left(1+\frac{w^2}{2}\right) w_i\,,
\end{align}
where we have expanded to $\oo(2)$ assuming $w^i = \oo(0.5)$ following
standard PPN counting. Once the solution is obtained, one performs a 
gauge transformation including terms involving $w$ to obtain the metric
in PPN form, \refeqs{g00ppn}{gijppn}. One can immediately see why this theory predicts preferred frame effects: $\dph(t)$ is only isotropic in 
the CMB rest frame. In any other frame it has a spatial gradient proportional to $w_i$. In any scalar-tensor theory, an evolving background field 
leads to
preferred-frame effects. In purely conformal theories, these
are suppressed by powers of the ratio of the Solar System time-scale (years) to the Hubble time
since $\dph\sim H_0$. These effects are usually neglected and we do the same here. The disformal coupling, on the other hand, adds additional explicit
preferred-frame effects that scale as $\dph/\Lambda$ and thus are not suppressed if $\Lambda \sim H_0$. This is required to obtain novel 
effects in cosmology \cite{vandeBruck:2015ida}.}

The second approach solves the field equations in the rest frame of the 
CMB, where the bulk motion of the Solar System is included in the 
energy momentum tensor; to lowest order, $v^{\rm CMB}_i(\vec{x}) = v^{\rm SS}_i(\vec{x}) + w_i$. One then performs a gauge transformation to the 
PPN
gauge in the absence of bulk motion, i.e. \refeqs{g00ppn}{gijppn} with
$w_i$ set to zero. Finally, the PPN metric is boosted to the Solar System
frame by a Lorentz boost (``post-Galilean transformation''), as described
in Sec.~4.3 of \cite{will1993theory}, which reintroduces the terms involving $w^i$
in \refeqs{g00ppn}{gijppn}.

Both approaches are equivalent since the underlying theory is 
Lorentz invariant, and since the physical relative velocity between
CMB and Solar System frames is included in either case (it is merely
absorbed in $\tilde{T}_{\mu\nu}$ in the second approach). While
the first approach in principle keeps the physics more clear,
the gauge transformation from the Einstein-frame, in which
the field solution is obtained, to PPN gauge is somewhat cumbersome in this
approach, as many terms
involving powers of $w^i$ and various contractions with PPN potentials
need to be kept. The second approach on the other hands avoids these
complications and so we follow it here. Thus, we begin by deriving the Einstein frame field solution in the CMB rest-frame, and then transform the 
resulting CMB Jordan-frame metric to the PPN form with $w=0$.

\subsubsection{Ansatz for the Solution}

We expand the field as
\begin{equation}
 \phi=\ph +\phh +\phhh,
\end{equation}
where $\ph$ is the cosmological background field in the CMB frame, while $\phh\sim\oo(1)$ and $\phhh\sim\oo(2)$. 
We expect that $\dph\sim H_0$ and $\ddph\sim H_0^2$ in the CMB frame since the evolution of $\ph$ is driven by the cosmological background (recall 
that $\phi$ is dimensionless in our convention). 
Looking at equation (\ref{eq:phi_eom}) with $\alpha_\phi=\bpp=0$ one can see that, at lowest order, there are two parameters multiplying the source terms 
for the field:
\begin{align}
\upo \equiv \frac{ \dph^2}{\Lambda^2} \quad\textrm{and}\quad
\upt \equiv \frac{ \ddph}{\Lambda^2}.
\end{align}
We will take both of these to be small numbers and will work only to leading-order in both. We will see that this approximation is 
self-consistent 
once the constraints from the PPN parameters have been imposed. We expand the Einstein frame metric as
$g_{\mu\nu}=\bar{g}_\nm+ h_\nm$, and impose the gauge choice
\begin{align} \label{gauge}
\partial_\mu h^\mu_0-\frac{1}{2}\partial_0h^\mu_\mu&=-{ \frac{1}{2}}\partial_0h_{00}\quad\textrm{and}\\
 \partial_\mu h^\mu_i-\frac{1}{2}\partial_ih^\mu_\mu&=0,
\end{align}
with $h^\mu_\nu\equiv \eta^{\mu\alpha} h_{\alpha\nu}$. 
With this gauge choice, one can write the Einstein frame metric as\footnote{Note that the symbol $\chi$ is often used to denote a quantity referred 
to as the \emph{superpotential} in the literature \cite{will1993theory}. We will not use the superpotential in this work and use $\chi$ to 
refer to perturbations of the $00$-component of the metric.}
\begin{equation}
\begin{array}{l l}
 g_{00}=-1+2\chi_1+2\chi_2,&\quad g^{00}=-1-2\chi_1-2\chi_2-4\chi_1^2\\
 g_{0i}=B_i,&\quad g^{0i}={ -}B^i\\
 g_{ij}= (1+2\Psi_1 )\delta_{ij},&\quad g^{ ij}=\left(1-2\Psi_1{ +}4\Psi_1^2\right){ \delta^{ij}},
\end{array}
\end{equation}
where our gauge choice implies
\begin{align}
 \partial_kB^k&=3\partial_0\Psi_1\label{eq:gauge1}\\
 \partial_0B_k&=\partial_k\chi_2,\label{eq:gauge2}.
\end{align}
The PPN order of the metric perturbations is $\chi_1\sim\Psi_1\sim\oo(1)$, $B_i\sim\oo(1.5)$ and $\chi_2\sim\oo(2)$. 
The resulting expression for the Jordan frame metric to $\oo(2)$ is given below in \refeq{g00int2}.

\subsubsection{The Energy-Momentum Tensor}

As remarked above, the energy-momentum tensor must be defined in the Jordan frame and so one has
\begin{equation}\label{eq:EMJF}
 \tilde{T}^{\nm}=\rho\left[1+{ \frac{p}{\rho}}+\Pi\right] u^\mu u^\nu+p\tilde{g}^\nm,
\end{equation}
where $u^\mu=\dd x^\mu/\dd \tilde{s}$, $\tilde{s}$ being the proper time for an observer in the Jordan frame. The quantities are $\rho\sim\oo(1)$, the 
density, $p\sim\oo(2)$, the pressure and $\Pi\sim\oo(1)$, the specific internal energy per unit mass, in the rest frame of the fluid.

We will ultimately want to change coordinates so that the lapse is unity, which we can do by setting $\dd T= N\dd t$, and so the velocity measured 
by an observer in the Jordan frame is 
\begin{equation}
 v^i=\frac{\dd x^i}{\dd T}=\frac{1}{N}\frac{\dd x^i}{\dd t}.
\label{eq:vi}
\end{equation}
In this case we have 
\begin{equation}
 u^\mu=\frac{\dd x^\mu}{\dd \tilde{s}}=\left(\gti,\gti N v^i\right),
\label{eq:umu}
\end{equation}
where $\gti \equiv \dd t/\dd \tilde{s}$ and $\tilde{s}$ is the proper time for an observer. In this case, one can calculate $\gti$ using the 
normalisation condition 
$\tg_{\mu\nu} u^\mu u^\nu=-1$ to find
\begin{equation}
 \gti=\frac{1}{N}\left(1+\frac{\chi_1}{N^2}+\frac{v^2}{2}\right)+\oo(2),
\end{equation}
Using this, we can find the Jordan-frame energy-momentum tensor to $O(2)$:
\begin{align}
\tilde{T}^{00}&=\frac{\rs}{N^2}\left[1 +\Pi+{v^2}+\frac{2\chi_1}{N^2} \right]\\
 \tilde{T}^{0i}&=\frac{\rs v^i}{N}\\
 \tilde{T}^{ij}&={\rs v^i v^j}+p\delta^{ij} .
\label{eq:Tmnupper}
\end{align}
We can then find the Einstein-frame energy-momentum tensor using equation (\ref{eq:emtrel}). Note that $\sqrt{1-2X/\Lambda^2} = 1-\upo/2 + \oo(1) = N 
+ \oo(1)$. Since $\tilde{T}^{0i}\sim\oo(1.5)$ and 
$\tilde{T}^{ij}\sim\oo(2)$ one simply has $T^{0i}=N\tilde{T}^{0i}$ and $T^{ij}=N\tilde{T}^{ij}$.  
$\tilde{T}^{00}$ contains both $\oo(1)$ and $\oo(2)$ 
terms. To $\oo(1)$ we have $X=(1 - 2\chi_1)\dph^2/2$ and hence
\begin{align}
%
{T}^{00}=\frac{\rs}{N}\left[1+\Pi+{v^2}+\frac{2\chi_1}{N} \right]. 
\end{align}
We will also need the lowered form of the energy-momentum tensor and the trace. One finds 
\begin{align}
 T_{00} &= \frac{\rs}{N}\left[1+\Pi+{v^2}
-2 \chi_1 N \right],\\
 T_{0i} &= -T^{0i} = -\rs v^i \\
 T_{ij} &= T^{ij} = N [ \rs v^i v^j + p\delta^{ij} ]\\
 T&= -\frac{\rs}{N}\left[1+\Pi+{v^2}+{ \chi_1} \upo \right]+\rs v^2 N+ 3pN.
\label{eq:Tmnlower}
\end{align}

These are all the quantities that we need to compute the solution at 2PN. We will do this by solving the trace-reversed form of Einstein's equation,
\begin{equation}\label{eq:TR}
 R_{\nm}=8\pi G\left(T_\nm -\frac{1}{2} g_\nm T\right),
\end{equation}
and the equation of motion (\ref{eq:phi_eom}) for $\phi$ to the appropriate order. Note that the contribution of the scalar field to 
the energy-momentum tensor contains terms proportional to $\dph$ that are unpaired with factors of $\Lambda^{-1}$, which we neglect. Compared to the 
matter variables and terms proportional to $\upo$ and $\upt$, they are suppressed by the ratio of the dynamical time of the system (of 
order a year in the Solar System) to the Hubble time $H^{-1}$ which is of order $10^{10}$ years.

\subsection{Solution at 1PN}

At 1PN the only quantities we need to calculate are $\chi_1$, $\Psi_1$ and $\phh$. We are working in the Einstein frame, where due to the absence of
anisotropic stress we have $\chi_1=\Psi_1$. The $00$-component of (\ref{eq:TR}) gives
\begin{equation}
 \nabla^2 \chi_1=\nabla^2\Psi_1=-\frac{4\pi G \rs}{N},
\end{equation}
the solution of which is
\begin{equation}\label{eq:c1p1sol}
 \chi_1=\Psi_1=\frac{U}{N}.
\end{equation}
Next, we need the field equation for $\phh$, which is
\begin{equation}
 \nabla^2\phh= 8\pi G \, \upt\, \rs ,
\end{equation}
which is solved by
\begin{equation}\label{eq:p1sol}
 \phh=-2\upt U
\end{equation}
to leading-order in $\upt$. These are all of the solutions at 1PN.

\subsection{Solution at 2PN}

At 2PN we need to find the metric potentials $\chi_2 \sim \oo(2)$ and $B_i \sim \oo(1.5)$ as well as the field $\phhh$. The following identities will be useful:\begin{align}
\label{ident:d0diU}
U_{,0i}&=-\frac{N}{2}\nabla^2(V_i-W_i)\\
\label{ident:diUdiU} 
U_{,i}U^{,i}&=\nabla^2\left(\frac{1}{2}U^2-\Phi_2\right).
\end{align}
Note that these differ from their usual form in the literature \cite{will1993theory} because we are working in a coordinate system with a non-trivial 
lapse. 
Specifically, for our Jordan-frame metric \refeq{JFmetric0}, the continuity 
equation becomes at lowest order
\be
N^{-1} \dot\rho + \partial_i(\rho v^i) = 0\,. 
\ee
The components $R_{00}$ and $R_{0i}$ of the Ricci tensor are given in many standard references (see \cite{will1993theory} for example) but the calculation implicitly uses the 
1PN solutions (\ref{eq:c1p1sol}) and the identities (\ref{ident:d0diU}) and (\ref{ident:diUdiU}) in the standard forms without the factors of $N$. 
The reader attempting to reproduce the calculations in this subsection should bear this in mind and, in particular, calculate these components 
explicitly.

We begin with the vector $B_i$. The $0i$-component of (\ref{eq:TR}) is 
\begin{equation}
 \frac{1}{2}\nabla^2 B_i + \frac{1}{2}\Psi_{1,0i} = 8\pi G\rs v_i.
\end{equation}

Using the solution (\ref{eq:c1p1sol}) and the identity (\ref{ident:d0diU}), one finds
\begin{equation}\label{eq:Bsol}
 B_i=-\frac{7}{2}V_i-\frac{1}{2}W_i.
\end{equation}
Next, we find $\chi_2$ using the $00$-component of (\ref{eq:TR}), which gives
\begin{align}\label{eq:chi_2sol}
\chi_2&= -\frac{ U^2}{N^2} 
+2\Phi_1
+\frac{2}{N^3}\Phi_2
+\frac{1}{N}\Phi_3
+3N\Phi_4.
\end{align}
Finally, we need the scalar to second order. Using the scalar's equation of motion (\ref{eq:phi_eom}), we find 
\begin{equation}\label{eq:phi2eq1}
 \nabla^2\phi_2=\ddot{\phi}_1
- \Sigma \left(8\nabla^2\Phi_2+2\nabla^2\Phi_3{ +}2\nabla^2\Phi_1 \right)
\end{equation}
at $\oo(2)$. The term proportional to $\ddphh$ can be dealt with by combining the gauge conditions (\ref{eq:gauge1}) and (\ref{eq:gauge2}) to find
\begin{equation}
 \frac{3}{N}\ddot{U}=\nabla^2\chi_2.
\end{equation}
Using this, one finds
\begin{equation}\label{eq:phi2sol} 
\phi_2= \Sigma \left(\frac{2}{3}U^2-\frac{{ 10}}{3}\Phi_1-\frac{28}{3}\Phi_2-\frac{8}{3}\Phi_3-2\Phi_4 \right).
\end{equation}
These are all of the 2PN quantities.

\subsection{The Jordan Frame Metric}

Now that we have all of the metric potentials and field solutions, we can calculate the Jordan frame metric using (\ref{eq:JFmetric0}). We continue 
to work to leading-order in $\upo$ and $\upt$ in what follows and we obtain
\begin{align}
\tg_{00}&=-N^2\left(1-\frac{2\chi_1}{N^2}-\frac{2\chi_2}{N^2}-\frac{2 \dph\dphh}{\Lambda^2} \right) + \oo(3)\label{eq:g00int2} \\
\tg_{0i}&=B_i+ \frac{\dph \phi_{1,i}}{\Lambda^2}+\frac{\dph \phi_{2,i}}{\Lambda^2} + \oo(2.5) \\
\tg_{ij}&=\left(1+2\Psi_1\right)\delta_{ij} +\oo(2).
\end{align}
This result is clearly not in PPN form: the lapse is not unity, and the last term in $\tg_{00}$ is $\oo(1.5)$. Also, there are $\oo(1)$ and $\oo(2)$ 
terms in the $0i$-component. We need to perform a gauge transformation to get this into the standard PPN gauge.

We begin by changing coordinates such that $\dd T=N\dd t$. This leaves the $ij$-components unchanged but the other components are
\begin{align}
\tg_{TT}&=-1+\frac{2\chi_1}{N^2}+\frac{2\chi_2}{N^2}+\frac{2 \dph\dphh}{\Lambda^2}\quad\textrm{and}\\
\tg_{Ti}&=\frac{B_i}{N}+\frac{\dph \phi_{1,i}}{\Lambda^2}+\frac{\dph \phi_{2,i}}{\Lambda^2}\,,
\end{align} 
thus eliminating the lapse. 

Next, we need to perform a post-Newtonian gauge transformation $x^\mu\rightarrow \tilde{x}^\mu$ to bring this into the PPN gauge. This is a 
second-order transformation and so we write $\tilde{x}^\mu=x^\mu-\xi_1^\mu-\xi_2^\mu$, where $\xi_n$ is $\oo(n)$ in the PPN counting 
scheme. One must then expand the 
formula
\begin{equation} 
 \hat{g}_\nm(\tilde{x}^\alpha)=\frac{\partial x^\sigma}{\partial\tilde{x}^\mu}\frac{\partial 
x^\lambda}{\partial\tilde{x}^\nu}\tilde{g}_{\sigma\lambda}(x^\alpha(\tilde { x } ^\rho)),
\end{equation}
where a hat denotes the new metric in the new coordinate system, to second order. The perturbations in the new gauge can be computed explicitly using 
the relations given in \cite{Bruni:1996im,Malik:2008im,Villa:2015ppa}. Note that since the new metric is written in terms of the new coordinates we do 
not need to expand the metric potentials separately. 

The $\oo(1.5)$ term in $\tg_{00}$ as well as the $\oo(1)$ term in $\tg_{0i}$
can be removed by choosing
\begin{align}
\xi_1^\mu = \frac{\dph \phh}{\Lambda^2}\left(1\,,\,\, \vec{0}\right)\,.
\end{align}
Note that in the transformation of the metric, $\xi_1^\mu$ always induces
terms of order $\upo$. Since we work to linear order in that parameter,
any terms $\oo(\xi_1^2)$ can be neglected. 

Finally, we can remove the $\oo(2)$ term in $\tg_{0i}$ by performing a
time shift at $\oo(2)$,
\begin{align}
\xi_2^T =\frac{\dph}{\Lambda^2}\left(2\phhh-4\upt U^2\right)\,, \quad \xi_{2\,i}=0\,.
\end{align}
Inserting the field solution, the metric then becomes
\begin{align}
\tg_{00}&=-1 + \frac{2U}{N^3} - \frac{2 U^2}{N^4} 
+\frac{4\Phi_1}{N^2}
+\frac{4\Phi_2}{N^5}
+\frac{2\Phi_3}{N^3}
+\frac{6\Phi_4}{N} +\oo(3)\label{eq:g00}\\
\tg_{0i}&=- \frac1{N}\left[\frac{7}{2}V_i + \frac{1}{2}W_i\right] +\oo(2.5)
\\
\tg_{ij}&=\left(1+\frac{2U}{N}\right)\delta_{ij} +\oo(2).
\end{align}

Comparing (\ref{eq:g00}) with (\ref{eq:g00ppn}), we can see that the metric is still not in PPN form, since the coefficient of $U$ in $\tilde g_{00}$ is not unity. The reason for this is that $G\equiv 
\mpl^2/8\pi$ is not equal to the locally measured value of Newton's constant $\GN$. We need to normalise the metric to PPN form and so we define
\begin{equation}\label{eq:GNN}
 \GN\equiv \frac{G}{N^3} = G\left[1+\frac{3}{2}\upo \right].
\end{equation}
This is the gravitational constant measured in the solar system. Next, we rescale every metric potential so that $\GN$ and not $G$ appears 
in their definition. That is, $U$ is now defined such that 
$\nabla^2U=-4\pi \GN \rho$. Doing this, we find 
\begin{align}
\tg_{00}&=-1+{2U}-{2N^2 U^2} 
+{4N\Phi_1}
+{4N\Phi_2}
+{2\Phi_3}
+{6N^2\Phi_4}+\oo(3)\label{eq:g00F}\\
\tg_{0i}&=-\frac{7N^2}{2}V_i -\frac{N^2}{2}W_i +\oo(2.5)\\
\tg_{ij}&=\left(1+{2N^2 U}\right)\delta_{ij} +\oo(2).
\end{align}
This metric is in the proper PPN form in the CMB frame, where $w^i=0$. 
Next, we proceed to transform it to the Solar System frame.


\subsection{PPN metric in the Solar System Frame}

As above, we shall call the velocity of the Solar System's centre of mass relative to the CMB background $w^i$. The motion can be incorporated into our picture 
quite simply by performing a Lorentz transformation corresponding to $w^i$ on the result \refeq{g00F} in the CMB frame.  
This coordinate transformation preserves the post-Newtonian character of our metric 
since $w =\sqrt{w_iw^i}$ is the velocity with respect to the CMB dipole and hence $w_i\sim\oo(370 \textrm{ km/s}) \sim\oo(0.5)$). An explicit 
treatment of the transformation has been well-documented (Sec.~4.3 in \cite{will1993theory}) and our metric takes the form
\begin{align}
\tg_{00}&=-1+{2U}-{2N^2 U^2}+{4N\Phi_1}+{4N\Phi_2}+{2\Phi_3}
+{6N^2\Phi_4}+3\upo w^2 U + \upo w^i w^j U_{ij}+4 \upo w^i V_i \label{eq:g00SS},\\
\tg_{0i}&= -\frac{7N^2}{2}V_i -\frac{N^2}{2}W_i +\upo w_i U +\upo w^j U_{ij}, \\
\tg_{ij}&=\left(1+{2N^2 U}\right)\delta_{ij} .
\end{align}

At first glance, it might seem surprising that we were able to derive the preferred-frame effects by working in the CMB rest frame.  However, this is indeed
possible because preferred-frame effects are present implicitly within the matter velocities $v^i$ in the CMB-frame metric. To see 
this, note that at lowest order $v^i$ can be decomposed into 
\begin{align}
v^i =& v^i_{\rm SS} + w^i\,,
\end{align}
where $v^i_{\rm SS}$ is the matter velocity with respect to the centre of mass of the Solar System frame. Furthermore, notice that there are 
$\alpha_i$ contributions in (and only in) the coefficients of the PPN potentials whose definitions involve $v^i$.\footnote{With the exception of the 
potential $\mathcal{A}$, which does not appear in conservative theories without preferred-location effects.}


\subsection{The PPN Parameters}

We are finally in a position to extract the PPN parameters, which can be done by comparing the metric \refeq{g00SS} with equations 
(\ref{eq:g00ppn})--(\ref{eq:gijppn}).  We obtain the following non-vanishing PPN parameters:
\begin{equation}\label{eq:PPNparams}
 \gamma=\beta=1-\upo,\quad\alpha_1=-4\upo, \quad\textrm{and}\quad\alpha_2=-\upo.
\end{equation}
The parameters $\gamma$ and $\beta$ are the PPN parameters that are commonly modified in scalar-tensor theories. $\alpha_1$ and $\alpha_2$ 
parametrize preferred-frame effects due to the motion of the Solar System relative to the cosmological background gradient $\dph$. Due to the 
relative motion of the Solar System with respect to the cosmological rest frame, it induces (apparent) preferred spatial directions locally. We will 
discuss some observable consequences of this below.  

The preferred-frame parameter $\alpha_3$ as well as the integral conservation-law parameters $\zeta_i$ on the other 
hand all vanish, as does the Whitehead parameter $\xi$. One may worry 
that this is true to leading-order in $\upo$ and $\upt$ only and that one should go to next-to-leading-order to derive their values. This is not the 
case. According to a 
theorem of Lee, Lightman and Ni \cite{Lee:1974nq}, any diffeomorphism-invariant theory of gravity that can be derived from a Lagrangian is at 
least semi-conservative, which implies that the above mentioned parameters are zero to all orders.  Our theory falls into this class and so there is 
no need to go beyond leading-order. The Whitehead parameter is also zero to all orders since the Whitehead potential 
$\Phi_W$ does not appear in the solutions for the metric potentials or the field. Equation (\ref{eq:PPNparams}) constitutes the main result of this 
section. 

In appendix \ref{sec:gencalc} we show that the disformal contributions are unchanged
when considering the general conformal/disformal theory $A(\phi),\,B(\phi)$.  
Specifically, the conformal factor $\alpha_\phi \equiv \rm{d}\ln A(\phi)/\rm{d}\ln\phi$ is independently constrained
by the light-bending parameter $\gamma$ to be $\alpha_\phi \lesssim \oo(10^{-3})$.  
This reduces any interactions of conformal and disformal effects to 
negligible levels. The parameter values given in equation~(\ref{eq:PPNparams}) then remain valid apart from a trivial rescaling $\upo \to B^2(\phi_0)\upo$. 
The parameters $\beta_\phi \equiv \rm{d}\ln B/\rm{d}\ln\phi$ and $\bpp^{\prime}$ are only constrained very weakly ($\bpp\lsim\oo(10^{5})$, 
$\bpp^{\prime}\lsim\oo(10^{9})$) through their contributions to the PPN parameter $\beta$ (see appendix A).

\section{Constraints}
\label{sec:constraints}

\subsection{A concrete model}

We have seen above that only $\upo$ and not $\Sigma$ is constrained by solar systems experiments. The strongest constraint comes from $\alpha_2$, 
which is constrained to 
satisfy $|\alpha_2|<4\times 10^{-7}$ \cite{Nordtvedt,Will:2008dya} from limits on the Sun's spin precession, implying $|\upo| < 4 \times 10^{-7}$ (we will discuss this constraint in Section~\ref{sec:concl}).
In order to investigate the implications for disformal gravity, we study a concrete representative of the disformal models considered in the cosmological context:
\begin{equation}
 A(\phi)=e^{\alpha_\phi\phi},\quad B(\phi)=e^{\bpp (\phi-\phi_0)}\quad\textrm{and}\quad V(\phi)=m_0^2e^{-\lambda\phi}.
\end{equation}
Following the discussion above, we set the conformal coupling $\alpha_\phi=0$ from here on. This is the model studied by \cite{Sakstein:2014aca} who found a dark 
energy dominated fixed point with 
\begin{equation}\label{eq:fps}
 \dph=\lambda H_0\quad\textrm{and}\quad \ddph=-\frac{\lambda^3 H_0^2}{2}
\end{equation}
when $\lambda<\sqrt{6}$. Note that this implies $\upo=-2\upt/\lambda$, which is consistent with our assumption that $\upo\sim\upt$. Assuming that we 
are close to this fixed point today, the constraint becomes
\begin{equation}\label{eq:consl}
 \lambda^2 \left(\frac{H_0^2}{\Lambda^2}\right)^2<4\times10^{-7}.
\end{equation}
The region in the $\Lambda/H_0$--$\lambda$ plane where this is satisfied is shown below in figure \ref{fig:cons}. One can see that 
$\Lambda/H_0\gsim3\times 10^{3}$ is required, which gives $\Lambda\gsim 10^{-30}$ eV. The disformal coupling is often stated in the form 
\cite{Brax:2014vva}
\begin{equation}
 \mathcal{L}\supset\frac{ T^{\nm}\partial_\mu\phi\partial_\nu\phi}{\mathcal{M}^4}
\end{equation}
in the decoupling limit. In this case, one has the relation $\mathcal{M}^4=\mpl^2\Lambda^2$ \cite{Sakstein:2014isa} and so our constraint translates 
into the bound $\mathcal{M}\gsim 10^{2}$ eV. Note that this is two orders-of-magnitude stronger than the previous bound using solar system tests 
$\mathcal{M}\gsim \mathcal{O}(\textrm{eV})$ \cite{Sakstein:2014isa}, which was found using an estimate of the Eddington light bending parameter 
$\gamma$.

\begin{figure}[ht]\centering
\subfigure{\includegraphics[width=0.4\textwidth,height=0.42\textwidth]{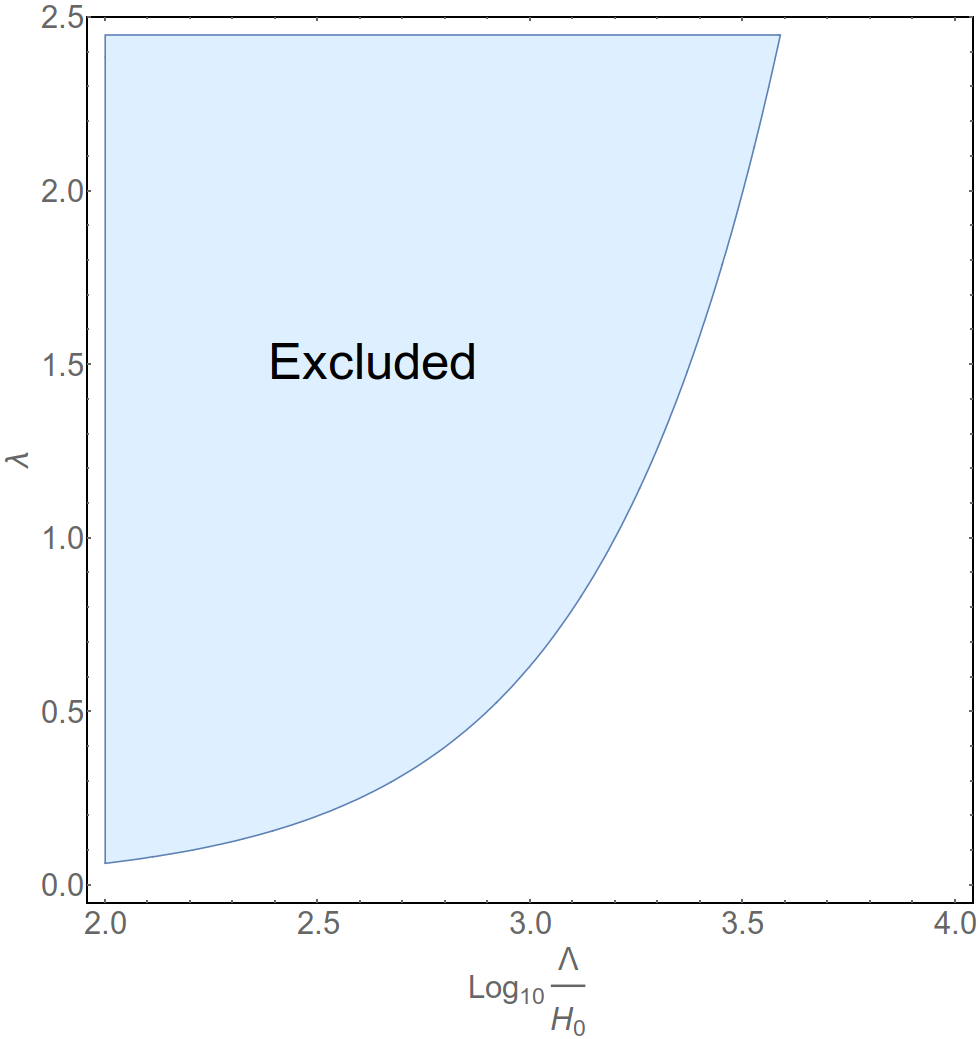}\label{fig:cons}}
\subfigure{\raisebox{4.4mm}{\includegraphics[width=0.4\textwidth,height=0.39\textwidth]{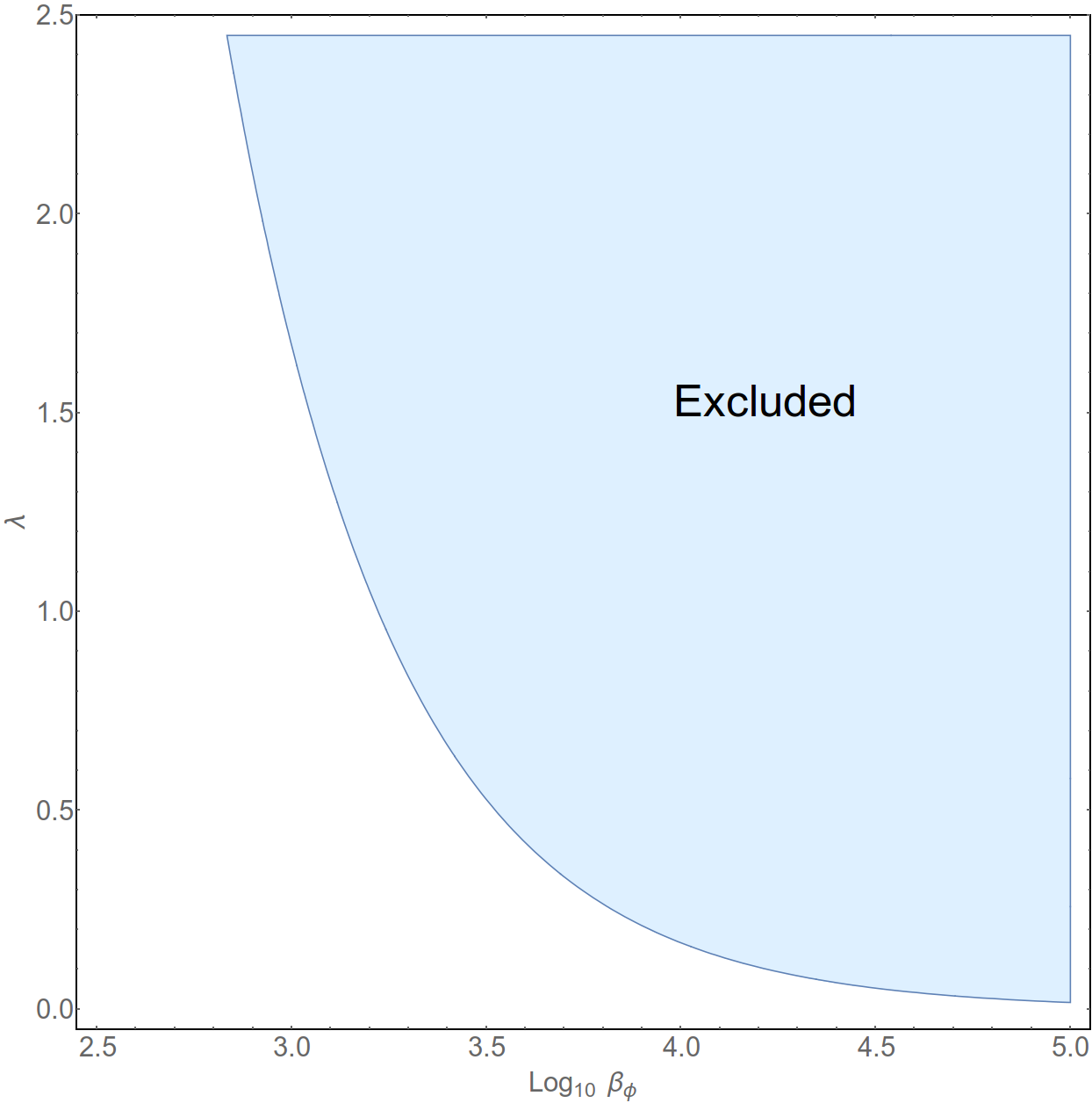}\label{fig:cons2}}}
\caption{{\bf Left:} The region in the $\Lambda/H_0$--$\lambda$ plane where the constraint from measurements of $\alpha_2$ is satisfied.\newline{\bf 
Right:} The region in the $\log_{10}\bpp$--$\lambda$ plane where the constraint from LLR measurements of the time-variation of $G$ are 
satisfied.}\label{fig:cons3}
\end{figure}

An independent constraint can be obtained from the time-variation of Newton's constant in the Solar System. Taking the time derivative of equation 
(\ref{eq:GNN}), we find
\begin{equation}
 \frac{\dot{G}_{\rm N}}{\GN}=\frac{3\dot{\upo}}{2(1+\frac{3\upo}{2})}\approx \frac{3\dot{\upo}}{2}.
\end{equation}
Lunar Laser Ranging (LLR) constrains this quantity to be less than $1.5\times10^{-13}$ yr$^{-1}$ \cite{Turyshev:2006gm}, which gives
\begin{equation}
\frac{3}{2}\lambda\Upsilon\left(2\bpp-\lambda\right)<0.002,
\end{equation}
where we have used equation (\ref{eq:fps}). This allows us to constrain $\bpp$ as a function of $\lambda$ by simultaneously imposing the constraint 
(\ref{eq:consl}). In figure \ref{fig:cons2} we plot the excluded region in the $\bpp$--$\lambda$ plane when $\Upsilon$ just satisfies the constraint 
(\ref{eq:consl}) i.e. $\upo=4\times10^{-7}$. One can see that $\bpp$ is relatively unconstrained and values as large as $10^3$ are not excluded.

\subsection{Implications for cosmology}

In this section we will show that the constraint $\frac{\Lambda}{H_0} \gtrsim 3\times 10^3$ implies that disformal effects are negligible in the 
context of the cosmological evolution. We begin by studying the expansion history 
described in the disformal model by the modified Friedmann equation in the Einstein frame \cite{vandeBruck:2015ida}
\begin{equation}\label{eq:fried}
H^2 = \frac{1}{3\mpl^2} (\rho_m+ \rho_\tn{de}),
\end{equation}
where\footnote{Comparing the definitions of the actions, $\phi=\frac{\sqrt{2}}{\mpl}\phi_{\tn{ref}}$ and $V=\frac{2}{\mpl^2}V_{\tn{ref}}$, where 
$\phi_{\tn{ref}}, V_{\tn{ref}}$ are the corresponding quantities defined in \cite{vandeBruck:2015ida}. Also, $D(\phi)={B^2}/{\Lambda^2\mpl^2}$ in our 
notation. }
\begin{align}
\dot{\rho}_m =&-3H \rho_m -\frac{Q\mpl\dot{\phi}}{\sqrt{2}},\quad
\rho_\tn{de} = \frac{\mpl^2}{2}\left(\frac{1}{2}\dot{\phi}^2+m_0^2e^{-\lambda\phi}\right)\quad\textrm{and}
\quad\ddot{\phi} + 2{H}\dot{\phi} - \lambda m_0^2e^{-\lambda\phi} = \frac{\sqrt{2}}{\mpl}Q \,.
\end{align}
Note that we have coupled the field to all species of matter, unlike
\cite{vandeBruck:2015ida} who only couple $\phi$ to the cold dark matter
component. The disformal effects enter through the coupling $Q$ given by
\begin{align}
\mpl Q=&   \frac{\rho_c}{\Lambda^2} \frac{\frac{3}{\sqrt{2}} H \dot{\phi}-\frac{\lambda m_0^2}{\sqrt{2}} e^{-\lambda \phi}-\bpp 
\dot{\phi}^2}{1+\frac{1}{\Lambda^2\mpl^2}\left( {\rho_m} -\frac{\mpl^2}{2}\dot{\phi}^2\right)}
\lsim 10^{-7}\,\rho_m   \quad(z=0),
\end{align}
where we have defined $\phi_0$ such that $B(\phi_0)=1$ and for the order-of-magnitude estimate at the epoch today, we have used
$\rho_m \sim \mpl^2 H_0^2$, $\dot\phi \sim H_0\phi$, and
$\Lambda/H_0 \gtrsim 3\times 10^3$. Note that the $10^{-7}$ bound was found assuming that $\beta_\phi\sim\oo(1)$, which is the theoretically natural 
value. This parameter is only weakly constrained by PPN measurements and we have found that it can be as large as $\oo(10^{3})$, in which case the 
bound above is weakened to $\mpl Q\lsim  10^{-4}\rho_m $. In fact, values of $\beta_\phi\gsim\oo(1)$ lead to phantom universes and are hence strongly 
disfavoured \cite{Sakstein:2015jca}. It is then clear that the fractional modification to the background expansion
\refeq{fried} due to the disformal coupling is of order $10^{-7}$--$10^{-4}$ or less for models that satisfy Solar System constraints. 
This is much smaller than current and near future observational uncertainties \cite{Amendola:2012ys}.
%

We now turn to the growth of linear density perturbations, $\delta_m = \delta \rho_m/\rho_m$.  
The growth equation for $\delta_m$ has been derived in \cite{vandeBruck:2015ida}\footnote{Note that \cite{vandeBruck:2015ida} work in conformal 
time. We have translated their results into coordinate time.}:
\begin{equation}\label{eq:growthfac}
\ddot{\delta}_m + \left(H -\frac{Q\mpl\dot{\phi}}{\sqrt{2}\rho_m}\right)\dot{\delta}_m= \left(4\pi G +\frac{Q^2}{\rho_m^2}\right) \rho_m \delta_m\,.
\end{equation}
This again reduces to the GR result if $Q$ is set to zero.  We then see
that the fractional correction to the growth factor is
constrained to be less than $10^{-7}$--$10^{-4}$ due to our by Solar System constraints.  
In fact, the effect will be smaller numerically, as the disformal coupling
only becomes relevant at late times, when $\rho_{\rm de}\sim \rho_m$, so that
the growth is modified by order $Q$ only over the last Hubble time.  

We see that for models satisfying the constraints from the Solar System
derived in the previous section, in particular from $\alpha_2$, the
impact on all cosmological observables is highly suppressed.  
Finally, we point out that there are two ways to evade these constraints:  
first, one could add one of the well-known screening mechanisms to the
scalar field action, i.e. the Chameleon or Vainshtein screening.  
Second, one could drop the weak equivalence principle and let the
scalar field couple only to dark matter and not to baryons.  In the
latter case, the Solar System constraints do not apply as they were derived
from baryonic objects.

\section{Conclusions}
\label{sec:concl}

In this work we have studied the local behaviour of disformal gravity theories to post-Newtonian order and have calculated the parameters appearing 
in the parametrised post-Newtonian (PPN) metric.  
The theory is semi-conservative and the Eddington light bending parameter $\gamma$, the non-linearity 
parameter $\beta$ and the preferred frame parameters $\alpha_1$ and $\alpha_2$ all differ from GR, with the differences proportional to $\upo \equiv 
\dph^2/\Lambda^2$. The strongest bound 
comes from $\alpha_2$, which imposes the constraint $\Lambda/H_0\gsim 3\times10^{3}$. This is two orders-of-magnitude stronger than the previous 
bounds using 
solar system constraints \cite{Sakstein:2014isa}. We were able to place a weak constraint on the first derivative of the disformal factor $\bpp\lsim 
10^3$ using Lunar Laser Ranging constraints on the time-variation of $\GN$. 

The constraint on $\alpha_2$ comes from the near perfect alignment of the Sun's spin axis with the orbital angular momenta of the planets. The term 
in the PPN metric proportional to $ \alpha_2$ leads to a torque on the Sun which induces a precession of the Sun's spin axis, contributing to the 
misalignment of the axes of spin and planetary orbital angular momenta. Ref.~\cite{Nordtvedt} obtained a constraint on $\alpha_2$ by integrating the 
motion of the Solar System relative to the cosmological reference frame over the past 5~Gyr.  
Caveats to this treatment given the uncertainties of galactic evolution have been pointed out in \cite{Iorio}.  As shown there, the orbit of Mercury 
provides an independent constraint $|\alpha_2| \lesssim 4\times 10^{-5}$, with further improvements possible.  While a detailed 
discussion of the $\alpha_2$ constraints goes beyond the scope of this paper, we point out that an integration time of 5~Gyr is still relatively short 
compared to the Hubble time, over which the background scalar field evolves.

In terms of the cosmology, deviations from GR are non-negligible at the background and linear level when $\mathcal{M}\lsim\mathcal{O}(10^{-3}\ \textrm{eV})$ 
\cite{vandeBruck:2015ida} and here we have constrained $\mathcal{M}\gsim 10^{2}$ eV.  In agreement with \cite{Sakstein:2014isa}, we have found that 
there is no ``disformal screening'' in these theories. These constraints then rule out disformal theories as a 
potential driving mechanism of the cosmic acceleration since they require a cosmology that is indistinguishable from $\Lambda$CDM. In this sense, the 
acceleration is due to the cosmological constant. One caveat is that the theory we have considered here assumes that there is a Jordan frame and so 
the scalar couples universally to all matter species. For this reason, the theory satisfies the equivalence principle. If one were to break this 
assumption and couple to dark matter only our constraints would be circumvented since solar system objects would not source the field.

We end by discussing the prospects for improving our constraints using other astrophysical probes. Binary pulsars have provided some of the most 
stringent tests of general relativity and conformal scalar-tensor theories to date \cite{Damour:1992we}, and so one may wonder whether the same is 
true for disformal theories. In addition to the PPN formalism, there is a parametrised post-Keplerian (PPK) framework for binary pulsar 
observations. Ref.~\cite{Sampson:2013wia} have shown that three of the PPK parameters can be obtained directly from the PPN parameters $\gamma$ and 
$\beta$. These are constrained to the $10^{-6}$ level at most\footnote{These bounds come from the binary pulsar system PSR J0737-3039 
\cite{Kramer:2006nb}.} and so one cannot improve the constraints found here using these measurements. The most accurately measured parameter is 
$P^{\rm PPK}$, the rate of orbital decay. This is constrained to the $10^{-12}$ level, however, the power emitted into scalar radiation typically 
scales as the square of the scalar charge, which scales like $\upo^2\,,\Sigma^2$ (see \cite{Sakstein:2014isa}) and so one expects this to yield 
constraints at the $10^{-6}$ level, which are not as strong as the bound obtained using $\alpha_2$, although a more detailed calculation is required 
to confirm this. Future observations that constrain the PPK parameters to higher 
precision have the potential to improve our bounds, but, for now, they are the strongest that one can obtain using the properties of slow-moving astrophysical objects alone.

\section{Acknowledgements}

We are grateful for several enlightening discussion with Eleonora Villa and Marco Bruni.

\appendix

\section{Calculation of the PPN Parameters: General Theory}
\label{sec:gencalc}
In this appendix we will generalise our calculation in section \ref{sec:mincase} to the general theory where the Jordan frame metric is
\begin{equation}
 \tg_\nm=A^2(\phi)\left(g_\nm+\frac{B^2(\phi)}{\Lambda^2}\partial_\mu\phi\partial_\nu\phi\right).
\end{equation}
Now that we have gained some intuition from the previous calculation, it is possible to greatly simplify this without the need to repeat the entire 
calculation using brute force. Recall from the previous section that, even though we calculated the field to 2PN, neither the 1PN nor 2PN 
field contributed to the PPN parameters at leading-order. The reason for this was that $\phi_{1,2}\sim \Sigma$ and their contribution to $\tg_{00}$ 
scaled like $\Sigma\phi_{1,2}$ which meant they only contributed terms that were higher-order in $\Sigma$. We therefore examine the changes to the 
calculation that occur in the general case to discern whether there are any new leading-order contributions to the Jordan frame metric. Purely 
conformal contributions are already constrained by previous analyses of scalar-tensor theories \cite{EspositoFarese:2004cc} and so we are interested 
to see if pure disformal and mixed terms will yield new constraints after the calculation. 

There are two new parameters that enter in the general 
case: $\alpha_\phi\equiv\alpha_\phi(\phi_0)$ and $\bpp\equiv\bpp(\phi_0)$, defined in equation~(\ref{eq:agdefs}).  
Expanding in these, one has, to $\oo(2)$
\begin{align}
A(\phi) &= A^2\left[1+ 2\phh \alpha_\phi + \left(2\phh^2 \alpha_\phi^2 +2\phhh \alpha_\phi+\phh^2 \alpha_\phi'\right)\right],\quad\textrm{and}\\
B(\phi) &= 1+ 2\phh \bpp + 2\phh^2 \bpp^2 +2\phhh \bpp+\phh^2 \bpp',
\end{align}
where $A^2\equiv A(\phi_0)^2$ and we have set $B(\ph)=1$ as before since we can absorb it into $\Lambda$. In addition to this, the definition of $\upo$ and $\upt$ are modified to include a factor of  $B^2(\ph)$ so 
that $\upo\rightarrow B^2(\ph)\upo$ and $\Sigma\rightarrow B^2(\ph)\Sigma$. In fact, since we are setting $B(\phi_0)=1$ this is not important in what 
follows but it does have implications for the LLR constraints derived in section \ref{sec:constraints} where we take time-derivatives of $\upo$. 
Using the same Einstein frame coordinates as the previous calculation, the Jordan frame metric is
\begin{align}
\tg_{00}&= 
-N^2A^2\left[1-2\frac{\chi_1}{N^2}-2\frac{\chi_2}{N^2}-2\bpp\frac{\upo}{N^2}\phh-\frac{\upo}{N^2}
\left(2\bpp\phhh+2\bpp^2\phh^2+\bpp^\prime\phh^2\right)\nonumber\right.\\&\left.-2\frac{\dph\dphh}{N^2\Lambda^2}
+2\alpha_\phi\phh+2\alpha_\phi\phhh+2\alpha_\phi^2\phh^2+\alpha_\phi^\prime\phh^2-4\alpha_\phi\frac{\phh\chi_1}{N^2}-4\alpha_\phi\bpp\frac{\upo}{N^2}\phh^2\right 
],\label{eq:jfint001}\\
\tg_{0i}&=A^2B_i+A^2\left[\frac{\dph}{\Lambda^2}\partial_i\phi_1+\frac{\dph}{\Lambda^2}\partial_i\phhh+2(\alpha_\phi+\bpp)\frac{\dph}{\Lambda^2}
\phh\partial_i\phh\right ]\quad \textrm{and}\\
\tg_{ij}&=A^2\left[1+2\alpha_\phi\phh+2\Psi_1\right]\label{eq:jfintij3}.
\end{align}
If we want to find the field to 1PN we only need the energy-momentum tensor (and its trace) to this order. At zeroth-order, one has $\gti=N^{-1}$, 
where the Jordan-frame Lorentz factor $\gti$ is defined below equation~(\ref{eq:umu}), and 
so one finds
\begin{align}
 T^{00}=-T=\frac{A^4\rho}{N},
\end{align}
where we have used the general relation between the two energy-momentum tensors
\begin{equation}\label{eq:emtrelgen}
T^{\nm}=A^6(\phi)\sqrt{1-\frac{2B^2(\phi)X}{\Lambda^2}}\tilde{T}^\nm.
\end{equation}
The scalar field equation at 1PN is then
\begin{align}
\nabla^2\phh=\frac{8\pi G\rs A^4}{N^3}f(\phi_0,\dph,\ddph)\quad\textrm{with}\quad f(\phi_0,\dph,\ddph)\equiv 
\alpha_\phi-\Upsilon(\alpha_\phi-\bpp)+\Sigma,
\end{align}
so that the 1PN solution is 
\begin{align}\label{eq:p1solgen}
 \phh = -2\frac{A^4 f}{N^3}U.
\end{align}
The general case differs from the minimal one in that it contains factors of $\upo$ and $\alpha_\phi$ multiplying $U$. In the purely conformal case 
where $\bpp=\upo=\upt=0$ one has $f=\alpha_\phi$ and the Eddington light bending parameter is \cite{EspositoFarese:2004cc}
\begin{equation}
 |\gamma-1|=\frac{2\alpha_\phi^2}{1+2\alpha_\phi^2}.
\end{equation}
One can see from (\ref{eq:jfint001})--(\ref{eq:jfintij3}) that this contribution is not affected by disformal contributions and so, in the absence of 
any fine-tuning\footnote{By this, we mean that we ignore cases where two or more parameters are fine-tuned against each other. One example of this is 
a cosmic evolutions such that $\frac{B(\ph )\dph}{\Lambda}\sim \Sigma$. \cite{Sakstein:2014aca} have shown that models where this is the case are 
unviable since they predict Newtonian limits that are not compatible with solar system, tests. The absence of fine-tuned models ensures that our 
counting scheme is self-consistent. }
or screening mechanism, the Cassini bound $|\gamma-1|<2.1\times10^{-5}$ \cite{Bertotti:2003rm} imposes the constraint $\alpha_\phi\lsim 10^{-3}$. The 
lapse $N$ is unchanged in the general case (this is a consequence of choosing $B(\phi_0)=1$) and so, looking at 
(\ref{eq:jfint001})--(\ref{eq:jfintij3})\footnote{One should really perform the generalised versions of the gauge transformations used in section 
\ref{sec:mincase}, but it is clear that these just add terms proportional to $\upo\phi_{1,2}$ and $(\alpha_\phi+\bpp)f^2U^2$ to $\tg_{00}$, which 
do not circumvent our arguments relating to (\ref{eq:jfint001})--(\ref{eq:jfintij3}).}, the only non-purely conformal leading-order corrections to 
the PPN parameters found in the minimal theory must be proportional to $\alpha_\phi\bpp\upo$ or $\alpha_\phi\Sigma$\footnote{$\alpha_\phi^\prime\lsim\oo(10^{2})$ 
from measurements of the perihelion shift of Mercury, which constrains the PPN parameter $\beta<3\times10^{-3}$ \cite{Will:2008dya}. Even taking the 
most extreme value does not alter the conclusions presented here.}. Demanding that we do not fine-tune different contributions to the PPN parameters 
means that $\Upsilon$ is still constrained by the $\alpha_2$ constraint such that $\upo,\,\,\upt\lsim 10^{-7}$. Therefore, any additional 
contributions present in the general theory automatically satisfy the PPN constraints. In particular, there are no new bounds on the conformal 
parameter $\alpha_\phi$ and the 
parameter $\bpp$ is completely unconstrained. The one caveat to this is that we have assumed that $\bpp\sim\oo(1)$. One can see that there is 
a very weak requirement that $\bpp\lsim 10^{5}$ due to a contribution to $\gamma$ of $\oo(\alpha_\phi\bpp\upo)$ and a similar constraint $\bpp\lsim 
10^{4.5}$ coming from a contribution to $\beta$ of $\oo(\alpha_\phi^2{\bpp}^2\Upsilon)$. Similarly, one can see that 
there is a contribution to $\beta$ of $\oo(\bpp^\prime\upo)$, which imposes the weak constraint $\bpp^\prime\lsim 10^{9}$. Note that these bounds 
apply when $\alpha_\phi$ and $\upo$ just satisfy their bounds i.e. $\alpha_\phi^2\sim \oo(10^{-5})$ and $\upo\sim \oo(10^{-7})$. When $\alpha_\phi$ and $\upo$ 
assume values smaller than this, $\bpp$ and ${\bpp}^\prime$ can assume larger values and still satisfy solar system tests.

\bibliographystyle{jhep}
\bibliography{ref}

\end{document}